# The Holographic Universe

Jean-Pierre Luminet

Aix-Marseille Université, CNRS, Laboratoire d'Astrophysique de Marseille (LAM) UMR 7326
& Centre de Physique Théorique de Marseille (CPT) CNRS-UMR 7332
& Observatoire de Paris (LUTH) CNRS-UMR 8102
France
E-mail: `jean-pierre.luminet@obspm.fr`

**Abstract**

I give a critical review of the holographic hypothesis, which posits that a universe with gravity can be described by a quantum field theory in fewer dimensions. I first recall how the idea originated from considerations on black hole thermodynamics and the so-called information paradox that arises when Hawking radiation is taken into account. String Quantum Gravity tried to solve the puzzle using the AdS/CFT correspondence, according to which a black hole in a 5-D anti de Sitter space is like a flat 4-D field of particles and radiation. Although such an interesting holographic property, also called gauge/gravity duality, has never been proved rigoroulsy, it has impulsed a number of research programs in fields as diverse as nuclear physics, condensed matter physics, general relativity and cosmology. I finally discuss the pros and cons of such an hypothesis, and emphasizes the key role played by black holes for understanding quantum gravity and the possible dualities between distant fields of theoretical physics.

**Introduction**

At a conference in Stockholm on August 28, 2015, Stephen Hawking announced that he had resolved the information paradox, a long-standing problem in theoretical physics.[1] The paradox suggests a potential conflict between quantum mechanics and the models of black holes described by general relativity. As such, it is a key aspect of fundamental physics and has divided the community of theoreticians for four decades.

According to Hawking, all the information about the matter and energy within the three-dimensional volume of a black hole is encoded as a hologram upon its two-dimensional surface, the event horizon.[2] This information may later be fully recovered, albeit in chaotic form, through the radiation released during quantum evaporation of the black hole, a process originally predicted by Hawking himself forty years ago.

The idea is not new. The holographic universe model has been studied by hundreds of physicists; some of the ensuing theories have been utterly surreal.[3] The scientific community reacted to Hawking's announcement with caution, skepticism, and, in some cases, even with embarrassment. It seemed a premature announcement for an idea not yet explained at a technical level; to date no details have been offered about how information is recorded in the event horizon or how it might be recovered.[‡]

---

[‡] After I completed this article a technical paper has finally been provided : Stephen Hawking, Malcom Perry and Andrew Strominger, "Soft Hair on Black Holes," (2016), arXiv:1601.00921.

To appreciate the situation more clearly, let us first consider the thermodynamics of black holes.

## Black Hole Thermodynamics

The 1970s were the golden age of classical general relativity. Physicists demonstrated, first, that the final state of a black hole in equilibrium depends on only three parameters: mass M, angular momentum J, and electric charge Q. This makes a black hole the simplest object in all of physics. It was demonstrated, in the second place, that the dynamics of interacting black holes could be summarized by four laws, which are analogous to the laws of thermodynamics.[4] In particular, the second of these laws stipulates that the area of a black hole can never decrease over time, suggesting a close connection between the area of a black hole and the entropy of a thermodynamic system.

In 1973, Jacob Bekenstein suggested that a black hole could have a well-defined entropy proportional to the area of its event horizon.[5] In classical general relativity, a black hole prevents any particle or form of radiation from escaping from its cosmic prison. For an external observer, when a material body crosses an event horizon all knowledge of its material properties is lost. Only the new values of M, J, and Q remain. As a result, a black hole swallows an enormous amount of information. The Bekenstein–Hawking formula describes the entropy that might be assigned to this information:

$$S = c^3 A/4\hbar G,$$

where A is the area of the event horizon, $c$ is the speed of light in a vacuum, $\hbar$ is the normalized Planck constant, and G is Newton's gravitational constant.[6]

In 1975, Hawking showed that the final state of a black hole, characterized by its three parameters M, J, and Q, was not stable when certain quantum effects close to the event horizon were taken into account.[7] A semi-classical analysis, in which matter but not the gravitational field is quantized, suggested that black holes should evaporate by emitting black body radiation.[8] Such radiation is characterized by a thermal spectrum temperature $T_H = g/2\pi$, where $g$ is the surface gravity of the black hole. This radiation carries mass, angular momentum, and electric charge, and therefore decreases the total energy of the black hole until it finally evaporates.

Hawking then pointed to a paradox. If a black hole can evaporate, a portion of the information it contains is lost forever. The information contained in thermal radiation emitted by a black hole is degraded; it does not recapitulate information about matter previously swallowed by the black hole. The irretrievable loss of information conflicts with one of the basic postulates of quantum mechanics. According to the Schrödinger equation, physical systems that change over time cannot create or destroy information, a property known as *unitarity*.

This apparent contradiction between general relativity and quantum mechanics is the information paradox.

## String Quantum Gravity

The information paradox reflects our current inability to develop a coherent theory of quantum gravity. The semi-classical Hawking approximation ceases to be valid when the black hole is so small that the radius of curvature at the event horizon reaches the Planck length, $10^{-33}$ cm. At this point, matter, energy, and the gravitational field must all be quantized. A complete description of how the black hole evaporates, and whether its information is partially or completely recovered, requires a comprehensive treatment of quantum gravity.

With the rise of quantum field theory, the techniques of quantum mechanics were successfully applied to physical objects such as electromagnetic fields.[9] This theory underlies the standard model of elementary particle physics, and accounts for electromagnetic, strong nuclear, and weak nuclear interactions. It also allows one to calculate probabilities using series perturbation methods.

Feynman diagrams describe the paths of point particles and their interactions. Each diagram represents one contribution to an ongoing process of interactions. When calculating these interactions, physicists first consider the strongest contributions and then the lesser, and so on, until they reach the desired accuracy. But this method only works if the contributions become negligible as more and more interactions are taken into account. When this is the case, the theory is weakly coupled, and calculations converge to finite physical values. If not, the theory is strongly coupled, and standard methods of particle physics fail. This is precisely what happens in the case of the graviton, thought to be the particle that mediates the gravitational field. The graviton creates mass-energy and interacts with itself. This interaction creates new gravitons. These in turn interact, and so on, until divergence. Perturbation methods thus fail to quantize gravity; this failure has led physicists to explore other avenues.

Among various theories of quantum gravity, it has been string theory that physicists have most thoroughly studied, despite promising alternatives such as loop quantum gravity and non-commutative geometry.[10] According to string theory, the fundamental constituents of matter—quarks, leptons and bosons—are not dimensionless point particles, but elongated, vibrating one-dimensional objects, or strings. These can be either open, with free ends, or closed, in a loop. Their modes of vibration and rotation are quantized and can be associated with particles of given mass and spin; every string corresponds to an infinite variety of particles. Interaction between particles is described in terms of the joining and the dividing of strings.

The advantages of this formulation are obvious. First, it marks the end of the particle zoo; all elementary particles can now be reduced to two families: closed strings and open strings. Second, by allowing for a minimal spatial scale, strings allow one to avoid the singularities otherwise inevitable in quantum theories of strongly coupled fields.

But there is a price to pay. Space-time is ordinarily described in terms of four dimensions: three spatial, and one temporal. To ensure the mathematical coherence of string theory, space-time must acquire an additional six spatial dimensions.[11]

Furthermore, integrating supersymmetry into string theory can be done in five different ways, making for five superstring theories, and none of them complete.[12] In some, all strings are necessarily closed on themselves, forming loops; in another, the strings are open and their ends are free.

In the early 1990s, theoreticians noted certain relations, called dualities, among the five superstring theories, and conjectured that each of them represents a particular case of a meta-theory, which they called M theory.[13] M theory thus functions in a space-time of eleven dimensions.[14]

The additional spatial dimensions of M theory gave rise to new fundamental objects which are called p-branes. Here, p is the number of spatial dimensions of the object in question. The complete system of branes forms a multidimensional space-time, the matrix. Strings are conceptualized as 1-branes and our usual three-dimensional space is a 3-brane.[15] The ends of open strings rest on the 3-branes, while the closed strings representing gravitons live in other dimensions.[16]

String theory and M theory, it should be noted, are still under construction. Physicists have formulated approximate equations for strings and branes, but not exact ones. Nor do they know how to calculate countless physical quantities from string equations.

But they do know that, under strong coupling, branes bend space-time enough to generate black branes, a generalization of black holes from classical general relativity. This is why, in the 1990s, the information paradox could be reconsidered, this time in the context of string quantum gravity.

## The Holographic Hypothesis

Reconsidering the information paradox was initially a matter of recovering the laws of classical thermodynamics for black holes. The entropy and temperature of a black hole had to be calculated, using statistical quantum mechanics, as a function of its area and surface gravity.

The task was not easy. As in thermodynamics, entropy measures the total number of internal microscopic states corresponding to a given external state of the black hole, defined by its three parameters M, J, and Q. The final degrees of freedom from which one calculates entropy must be made countable.

To evaluate the final information content of a material element—i.e., its thermodynamic entropy—one must rigorously know its fundamental constituents at the deepest structural level. In the standard model of particle physics, this information is encoded by quarks and leptons. But in string theory and M theory, quarks and leptons are just excited states of superstrings, which are in turn the most elementary constituents of the physical world.

In 1993, Gerard 't Hooft was the first to revisit Hawking's work on the thermodynamics of black holes in the context of string theory.[17] He calculated that the total number of degrees of freedom in the volume of space-time inside a black hole was proportional to the surface area of its horizon.[18] The two-dimensional surface of a black hole can be divided into fundamental quantum units, called Planck areas ($10^{-66}$ cm$^2$).

From the point of view of information, each bit in the form of a 0 or a 1 corresponds to four Planck areas, which allows one to find the Bekenstein–Hawking formula for entropy. For an external observer, information about the entropy of the black hole, once borne by the three-dimensional structure of the objects that have crossed the event horizon, seems lost. But on this view, the information is encoded on the two-dimensional surface of a black hole, like a hologram. Therefore, 't Hooft concluded, the information swallowed by a black hole could be completely restored during the process of quantum evaporation.

In the more general context of M theory, Andrew Strominger and Cumrun Vafa managed in 1996 to calculate the entropy of a charged, extremal black hole in five dimensions.[19] Considering a black hole as a string gas and counting the quantum states associated with string vibrations, they were able to find the Bekenstein–Hawking formula for the entropy of a black hole as a function of the area of its horizon, its temperature as a function of its surface gravity, and its Hawking radiation characteristics.[20]

In the wake of these encouraging results, numerous researchers succeeded in finding the microscopic entropy of somewhat more generalized black holes—i.e., those in four and five dimensions with electric charges and angular momentums that are close to extremal.[21] From then on, the evaporation of a black hole could then be seen as the emission of closed strings (gravitons) from a system of branes representing the black hole.

Finally, string theory seemed ready to resolve the information paradox, and to prove that Hawking radiation contains all information about the internal properties of a black hole. This proved too optimistic. It was quickly noticed that all control over calculations is lost as soon as one departs significantly from extremal conditions. Microstates cannot be calculated for either the simplest black hole, the Schwarzschild solution, or the most realistic astrophysical black hole, the Kerr solution.

Nevertheless, the idea that the amount of information remaining within a black hole depends on the area of its event horizon and not on its volume aroused extraordinarily keen interest. Extensively developed by Leonard Susskind, it has become known as the holographic principle, with hopes that it can be generalized for *any* physical system occupying a volume of space-time.[22]

According to this principle, all the physics within a volume, including gravitational phenomena, can be entirely described by another physical theory that operates only on the border of the volume.

One might suspect that the holographic principle does not apply to our universe as a whole. The universe, to the best of our knowledge, is a space-time with four macroscopic dimensions; if it were holographic, a set of alternative physical laws that apply only to its three-dimensional border would exist somewhere, and would be in some way equivalent to the usual four-dimensional physics.

What surface could function as the border of our space-time? The standard cosmological model is an open Friedmann–Lemaître solution of the field equation of general relativity, with spatial curvature close to zero and accelerated expansion. As opposed to a black hole, this model has no definite border at which to place such a hologram. In addition, the holographic limit on entropy deduced from black holes is destroyed in an expanding, spatially homogenous, and isotropic universe; the entropy of a region of space-time full of matter and radiation that is not collapsed into a black hole is actually proportional to its volume, not to the area of its border.

# Juan Maldacena's Conjecture

Facing these difficulties, physicists turned to simplified models of the universe to which the holographic principle could be applied. In 1997, Juan Maldacena offered a stunning solution to these problems, together with an audacious mathematical conjecture.[23] Maldacena's original publication has been cited more than ten thousand times, making it the most cited article in all the literature of theoretical physics.

Maldacena considered a black hole in a model of space-time with five macroscopic dimensions—the so-called anti-de Sitter space.[24] Such a universe is described by string theory and thus includes gravitation. He showed that the details of the phenomena taking place in such a universe were, indeed, entirely encoded in the behavior of certain quantum, non-gravitational fields taking place on the four-dimensional border of this universe.

How does this work? Known since 1917, a de Sitter space-time is an exact solution of the equations of ordinary general relativity. It is empty of matter, but includes a positive repulsive force called a cosmological constant. If one changes the sign of the cosmological constant, the repulsive force becomes attractive, and the model becomes what is called an anti-de Sitter space-time.[25] This then acquires a hyperbolic spatial geometry—negative curvature. Even if it is an infinite space-time, it possesses a well-defined edge. To represent this edge, one uses Henri Poincaré's representation of a hyperbolic disk, which, with the help of a conformal transformation that conserves angles but not distances, reduces the infinite to a finite distance.[26]

For five-dimensional anti-de Sitter space-time, abbreviated as $AdS_5$, the edge is four dimensional and, locally around each point, resembles a Poincaré–Minkowski space. It is precisely the model of flat space-time used in non-gravitational physics. This means that a black hole in the five-dimensional anti-de Sitter space-time is strictly equivalent to a field of particles and radiation existing in the flat, four-dimensional space-time of the border. This last description relies on well-understood quantum field theories analogous to the Yang–Mills fields used in quantum chromodynamics, the theory of strong interactions.

What is most interesting about this equivalence is that gravitational physics in $AdS_5$ is in a regime of strong coupling and therefore not treatable in perturbation theory, while the non-gravitational physics on the four dimensional border is a gauge theory with weak coupling, and thus is calculable. Since gauge theories are well defined, whatever the strength of the coupling, Maldacena conjectured that his description would apply generally, regardless of the coupling intensity on either side of the equivalence.

His conjecture can be summarized thus :

$$\textit{String theory on } AdS_5 \times S^5 \sim \textit{Gauge theory on the 4D border,}$$

where ~ is the symbol of the duality.[27]

A four dimensional gauge theory is a conformal field theory (CFT).[28] Maldacena's conjecture was renamed the AdS/CFT correspondence, or the gauge/gravity duality, in order to accentuate the fact that gravity in the context of string theory emerges from gauge theory, and vice versa.[29]

The AdS/CFT correspondence joins together two types of theories. Theories of quantum gravity, formulated in terms of string or M theory, apply to the interior of a particular five-dimensional space-time; conformal field theories apply to its four-dimensional border, where they describe the elementary particles.

The AdS/CFT correspondence is thus often described as a holographic duality, because a five-dimensional universe is recorded like a hologram on its four-dimensional border.[30]

This is a surprising result. There are generally fewer bits of information on the surface of a volume than within the volume itself. If everything is encoded in a hologram, certain correlations between particles prevent them from being entirely independent. Holography imposes a limit of its own on the entropy of interior particles. It does not turn out to be significant except for black holes, or for very high density plasmas, neither of which is accessible to experimental testing. Even an approximation of this duality would still be interesting. Physics is, after all, the art of clever approximation.

One benefit of Maldacena's work has been to resolve the information paradox, at least in the particular case of a black hole in $AdS_5$. This is, it turns out, equivalent to a hot plasma on the border, characterized by the Hawking temperature $T_H$ and described by a gauge theory. The plasma and black hole both have the same entropy. Furthermore, the plasma obeys the usual laws of quantum mechanics; in particular it evolves unitarily, which means the black hole evolves unitarily, and also obeys the principles of quantum mechanics.[31]

This result led Hawking to revise his position, and to announce in 2005 that the information paradox had been completely resolved by the AdS/CFT correspondence in favor of the conservation of information.[32]

## Dualities

Hundreds of researchers have explored the consequences of Maldacena's conjecture, hoping that the gauge/gravity duality in its most general form would establish a kind of practical dictionary matching the properties of a physical system in quantum gravity, described by string or M theory in a curved space of high dimensionality (the matrix), to those of another, simpler, physical system, quantitatively described by a gauge theory on the border of the matrix, namely, a flat space of lower dimensionality.[33]

The advantages are obvious. Certain very complex calculations in quantum gravity can be treated much more simply in the context of gauge theory. Recall the quantum evaporation of a black hole in $AdS_5$. The strong/weak duality allows one to explore complex aspects of nuclear physics and of condensed matter physics by translating them into string-theoretic terms.

The gauge/gravity duality shapes the ambitions of programs in three broad fields of physics: nuclear physics, especially the study of quark–gluon plasmas (AdS/quantum chromodynamics program); condensed matter physics, with the study of exotic states of matter (AdS/condensed matter theory program); and general relativity and cosmology, with the Kerr/CFT and de Sitter/CFT programs.

Quantum chromodynamics (QCD) is the theory of the strong interaction between quarks and gluons. This interaction has the particular property of being strongly coupled at low energies, which makes calculations exceedingly difficult, but explains why isolated quarks remain confined to the nuclei. At high energies, however, the interaction becomes weaker and weaker (a property called asymptotic freedom), in such a way that above a certain temperature, around $2\times10^{12}$K, quarks and gluons escape to form a quark–gluon plasma.

This phenomenon is currently produced in particle accelerators by colliding heavy ions. Still, it must have played an essential role in the primordial universe, where temperature conditions during the $10^{-11}$ seconds after the big bang were likely comparable.[34] Researchers have shown that an AdS/QCD duality can be used to understand certain aspects of the quark–gluon plasma.[35] These results have clarified some of the early work of 't Hooft, suggesting that calculations in quantum field theory under certain conditions resemble those in string theory.[36]

Applications of the gauge/gravity duality to the quark-gluon plasma initially aroused great enthusiasm. There is also room for skepticism.[37] First, QCD is certainly a gauge theory, but, as opposed to the Yang–Mills theory used in the AdS/CFT correspondence, it is neither supersymmetric nor conformal. There is no equivalence between the two, only similarities that in no way guarantee the relevance of calculations in the AdS/QCD duality.

Second, experimental condensed matter physics has discovered a certain number of exotic states of matter, such as superconductivity and superfluidity. These states are described using quantum field theory, but are hard to treat when strongly coupled. Dual theories describing weakly coupled states are easier to use. The AdS/CMT program has therefore permitted modeling the transition of a superfluid to an insulator, non-Fermi liquids (also called strange metals), and high temperature superconductors.[38] In these three areas, the AdS/CMT correspondence has furnished the best, and sometimes the only, tool for dealing with complex physical problems. As analytical and numerical methods have developed, progress has been rapid. Still, researchers like Philip Warren Anderson have recently expressed doubts about the effectiveness of this approach.[39]

As for the case of the CFT dualities, most black holes in the context of the AdS/CFT correspondence reside in $AdS_5$ and are thus physically unrealistic. The Kerr/CFT program, begun in 2009, has tried to show that the holographic duality can be used to understand certain astrophysical black holes. The results, however, only apply to Kerr extremal black holes, which have the highest possible angular momentum, and for which the event horizon disappears.[40] Still, the Kerr/CFT correspondence has been extended to black holes with a lower angular momentum, allowing one to find the Bekenstein–Hawking formula for any value of M and J.[41]

Despite hundreds of articles written on the subject, no one has ever demonstrated that the AdS/CFT correspondence, or a similar duality, could apply to our physical universe.[42] The five-dimensional $AdS_5$ model used in the gauge/gravity duality is empty, static, and has a negative cosmological constant. It has nothing in common with our four-dimensional universe, which is filled with matter and energy and has a positive cosmological constant.[43]

Some researchers have attempted a possible dS/CFT correspondence, linking quantum gravity in de Sitter space to a conformal theory of Euclidean fields.[44] Such a duality would be interesting from the point of view of cosmology; many scholars believe that, during inflation, the early universe was very much like a de Sitter space, and could return to resembling one in the distant future if the current rate of accelerating expansion becomes exponential.

This duality does not seem to work, if only because the boundary of a de Sitter space-time cannot be defined properly. All other attempts to describe our universe holographically have failed.

The gauge/gravity duality is of no use in rational cosmology.

## Black Holism

As we have seen, the AdS/CFT correspondence and holographic dualities have aroused immense enthusiasm in the string theory community. This constitutes, after all, normal scientific research. The phenomenon is still puzzling.[45] At a minimum, the holographic duality is an interesting tool for calculating fundamental physics. The dictionary the duality offers—between a world in flat space-time and a curved world with gravity—works in both directions. Some calculations are simpler with supergravity than in dual gauge theory.

Gauge/gravity duality has enhanced the stature of Albert Einstein's own theory. General relativity has certainly enjoyed remarkable success over the last century, providing a critical building block for the entire side of theoretical physics that deals with gravitation. It has driven a conceptual revolution in how we think about the nature of space and time. Anyone with a minimum of scientific culture today has heard of Einstein's theory.

But, despite its recognized elegance, general relativity has been used by only a small portion of the scientific community. This is not surprising. After all, general relativity seemed confined to cases of strongly curved space-time: compact stars, the big bang, gravitational waves. Its effects were utterly negligible at the scales at work in condensed matter physics and nuclear physics. Why should gravity play a role in the quantum world? Yet, over the last twenty years general relativity has finally penetrated the world of modern physics. Specialists in condensed matter, nuclear physics, fluid turbulence, and quantum information are actively interested in general relativity.

Why this dramatic turnaround?

As science progresses, the virtues of cross-fertilization between different areas of knowledge have become widely appreciated. But this is the result rather than the cause. The key factor has been the AdS/CFT correspondence. Thanks to general relativity and its string theory extensions, one can now describe phenomena that have nothing to do with gravity in strong fields.

On the other hand, the AdS/CFT correspondence has not been mathematically demonstrated. The holographic principle remains a conjecture.

Its degree of experimental verification is zero.

String theorists believe in it because their theory supports a specific version of holography, and, under certain important restrictions, black hole thermodynamics suggests it as well. But to conclude that it is a correct representation of nature is an enormous leap.

We still do not know if string theory is correct. Let us suppose it is. Different formulations of the holographic principle have been tested only in situations that do not correspond to our world. The ensuing equations describe possible worlds that are similar, but not identical to our own. Extant solutions have allowed us to test the principles of string theory in limiting cases and to show the consistency of the theory, but never in situations corresponding exactly to the world in which we live.

Suppose, on the other hand, that string theory proves false. What would become of the AdS/CFT correspondence? In loop quantum gravity, space-time emerges as the coarse-graining of fundamental structures made discrete, the atoms of space. The formation of a black hole and its ultimate evaporation are described by a unitary process that respects the laws of quantum mechanics.[46] Similarly, the black holes described in the approach to non-commutative geometry do not evaporate completely and therefore escape the information paradox.[47] The holographic conjecture has nonetheless improved the epistemological status of black holes.

Their status was not always so elevated. The first exact solution of general relativity describing the space-time of a black hole was discovered by Karl Schwarzschild in 1916. Until the 1950s, general relativity theorists were embarrassed by black holes because of their singularities. Later they seemed esoteric objects, hardly believable. Then, in the decades between 1960 and 1990, they became both relevant to astrophysics, and fascinating in their own right. As we have seen, black holes have proven key for understanding quantum gravity, and the deep dualities between distant fields of theoretical physics. Perhaps someday they will become ubiquitous, because they have become useful in the description of everyday systems.

**Notes and References**

1. Stephen Hawking, "The Information Paradox for Black Holes," (2015), arXiv:1509.01147.
2. A hologram is a photograph of a particular type which generates a three-dimensional image when it is suitably illuminated; all the information describing a three-dimensional scene is encoded in the pattern of light and dark areas inscribed on a two-dimensional film.
3. For example, Samir Mathur has proposed that, instead of being destroyed by the tidal forces of gravitation or by a quantum firewall, an astronaut falling into a black hole would simply be converted into a hologram, without noticing anything. Samir Mathur, "A Model with no Firewall," (2015), arXiv:1506.04342.
4. For an educational introduction, see, for example: Jean-Pierre Luminet, *Black Holes* (Cambridge, UK: Cambridge University Press, 1992), chs. 11–14.
5. Jacob Bekenstein, "Black holes and entropy," *Physical Review D* 7, no. 8 (1973): 2,333. The author, the origin of all discussions of the information paradox, died August 16, 2015, to nearly total media indifference, while any public statement by Stephen Hawking regularly makes headlines around the world.
6. In what follows I posit that $c = G = \hbar = 1$, as is common usage in quantum gravity.
7. Stephen Hawking, "Particle Creation by Black Holes," *Communications in Mathematical Physics* 43, no. 3 (1975): 199–220.
8. At least this should be the case with mini-black holes of low mass that might have been formed by the big bang; the phenomenon is utterly negligible for astrophysical black holes.
9. Michael Peskin and Daniel Schroeder, *An Introduction to Quantum Field Theory* (Boulder, CO: Westview Press, 1995).